\documentclass[12pt]{article}
\usepackage[english]{babel}
\usepackage{graphicx,color} % for figures
\usepackage{bm}       % bold math
\usepackage{amsmath}
\usepackage{amssymb}
\usepackage{caption}
\usepackage[affil-it]{authblk}

\DeclareMathAlphabet{\mathpzc}{OT1}{pzc}{m}{it}
\begin{document}

\title{Cluster Separability in\\ Relativistic Few Body Problems}
\author{N. Reichelt and W. Schweiger}
\affil{Institute of Physics, University of Graz, A-8010 Graz,
Austria}
\author{W.H. Klink}
\affil{Department of Physics and Astronomy, University of Iowa, Iowa City,
USA}

\maketitle

\noindent {\bf Abstract:}  A convenient framework for dealing with hadron structure and hadronic physics in the few-GeV energy range is relativistic quantum mechanics. Unlike relativistic quantum field theory, one deals with a fixed, or at least restricted number of degrees of freedom while maintaining relativistic invariance. For systems of interacting particles this is achieved by means of the, so called, \lq\lq Bakamjian-Thomas construction\rq\rq , which is a systematic procedure for implementing interaction terms in the generators of the Poincar\'e group such that their algebra is preserved. Doing relativistic quantum mechanics in this way one, however, faces a problem connected with the physical requirement of cluster separability as soon as one has more than two interacting particles. Cluster separability, or sometimes also termed \lq\lq macroscopic causality\rq\rq, is the property that if a system is subdivided into subsystems which are then separated by a sufficiently large spacelike distance, these subsystems should behave independently. In the present contribution we discuss the problem of cluster separability and sketch the procedure to resolve it.

\section{Introduction to relativistic quantum mechanics}
It is a widespread opinion that a relativistically invariant quantum theory of interacting particles has to be a (local) quantum field theory. Therefore we first have to specify what we mean by \lq\lq relativistic quantum mechanics\rq\rq . Relativistic quantum mechanics is based on a theorem by Bargmann which basically states that~\cite{Bargmann:54,Kei:91}:\\
{\em A quantum mechanical model formulated on a Hilbert space preserves probabilities in all inertial coordinate systems if and only if the correspondence between states in different inertial coordinate systems can be realized by a single-valued unitary representation of the covering group of the Poincar\'e group.}\\
According to this theorem one has succeeded in constructing a relativistically invariant quantum mechanical model, if one has found a representation of the (covering group of the) Poincar\'e group in terms of unitary operators on an appropriate Hilbert space. Equivalently one can also look for a representation of the generators of the Poincar\'e group in terms of self-adjoint operators acting on this Hilbert space. These self-adjoint operators should then satisfy the Poincar\'e algebra
\begin{eqnarray}\label{eq:PCalgebra}
&&[J^i,J^j]=\imath\, \epsilon^{ijk} J^k\, , \quad [K^i,K^j]=-\imath\, \epsilon^{ijk} J^k\, , \quad [J^i,K^j]=\imath\, \epsilon^{ijk} K^k\, ,\nonumber \\
 &&\left[P^\mu,P^\nu\right] =0\, ,\quad [K^i,P^0]=-\imath\, P^i\, ,\quad [J^i,P^0]=0\, ,\nonumber\\
 &&{[J^i,P^j]=\imath\,\epsilon^{ijk}P^k\, ,\quad \left[K^i,P^j\right]=-\imath\, \delta_{ij}\,P^0}\, .
\end{eqnarray}
$P^0$ and $P^i$ generate time and space translations, respectively, $J^i$ rotations and $K^i$ Lorentz boosts. From the last commutation relation it is quite obvious that, if $P^0$ contains interactions, $K^i$ or $P^j$ (or both) have to contain interactions too. The form of relativistic dynamics is then characterized by the interaction dependent generators. Dirac~\cite{Dirac:49} identified three prominent forms of relativistic dynamics, the {\em instant form} (interactions in $P^0$, $K^i$, $i=1,2,3$), the {\em front form} (interactions in $P^-=P^0-P^3$, $F^1=K^1-J^2$, $F^2=K^2+J^1$) and the {\em point form} (interactions in $P^\mu$, $i=0,1,2,3$). In what follows we will stick to the point form, where $P^\mu$, the generators of space-time translations, contain interactions and $\vec{J}$, $\vec{K}$, the generators of Lorentz transformations, are interaction free. The big advantage of this form is that boosts and the addition of angular momenta become simple.

For a single free particle and also for several free particles it is quite easy to find Hilbert-space representations of the Poincar\'e generators in terms self-adjoint operators that satisfy the algebra given in Eq.~(\ref{eq:PCalgebra}), but what about interacting systems? Local quantum field theories provide a relativistic invariant description of interacting systems, but then  one has to deal with a complicated many-body theory. It is less known that interacting representations of the Poincar\'e algebra can also be realized on an $N$-particle Hilbert space and one does not necessarily need a Fock space. A systematic procedure for implementing interactions in the Poincar\'e generators of an $N$-particle system such that the Poincar\'e algebra is preserved, has been suggest long ago by Bakamjian and Thomas~\cite{Bakamjian:53}. In the point form this procedure amounts to factorize the four-momentum operator of the interaction-free system into a four-velocity operator and a mass operator and add then interaction terms to the mass operator:
\begin{equation}\label{eq:BT}
{P^\mu}={M} { V^\mu_{\mathrm{free}}}= ({ M_{\mathrm{free}}}+{ M_{\mathrm{int}}})  { V^\mu_{\mathrm{free}}}\,.
\end{equation}
Since the mass operator is a Casimir operator of the Poincar\'e group, the constraints on the interaction terms that guarantee Poincar\'e invariance become simply that $M_{\mathrm{int}}$ should be a Lorentz scalar and that it should commute with $V^\mu_{\mathrm{free}}$, i.e. $[M_{\mathrm{int}}, V^\mu_{\mathrm{free}}]=0$. Remarkably, this kind of construction allows for instantaneous interactions (\lq\lq interactions at a distance\rq\rq ). Similar procedures can also be carried out in the instant and front forms of relativistic dynamics such that the physical equivalence of all three forms is guaranteed in the sense that the different descriptions are related by unitary transformations~\cite{Shatnii:78}.

A very convenient basis for representing Bakajian-Thomas (BT) type mass operators consists of velocity states
\begin{equation}
\vert \vec{v}; \vec{k}_1, \mu_1; \vec{k}_2, \mu_2; \dots; \vec{k}_N, \mu_N
\rangle \, ,\qquad \sum_{i=1}^N \,
\vec{k}_i = 0\, .
\end{equation}
These specify the state of an $N$-particle system by its overall velocity $\vec{v}$, the particle momenta $\vec{k}_i$ in the rest frame of the system and the spin projections $\mu_i$ of the individual particles. The physical momenta of the particles are then given by $\vec{p}_i=\overrightarrow{B(\vec{v}) k_i}$, where $B(\vec{v})$ is a canonical (rotationless) boost with the overall system velocity $\vec{v}$. Associated with this kind of boost is also the notion of \lq\lq canonical spin\rq\rq\ which fixes the spin projections $\mu_i$. $N$-particle velocity states, as introduced above, are eigenstates of the free $N$-particle velocity operator $V^\mu_{\mathrm{free}}$ and the free mass operator
\begin{equation}
M_{\mathrm{free}}\, \vert \vec{v};
\vec{k}_1, \mu_1; \vec{k}_{2}, \mu_{2};\dots\rangle = (\omega_1+\omega_2+\dots)\, \vert \vec{v};
\vec{k}_1, \mu_1; \vec{k}_{2}, \mu_{2};\dots\rangle\, ,
\end{equation}
with $\omega_i=\sqrt{m_i^2+\vec{k}_i^2}$. The overall velocity factors out in velocity-state matrix elements of BT-type mass operators,
\begin{eqnarray}
\langle \vec{v}^\prime; \vec{k}_1^\prime, \mu_1; \vec{k}_{2}^\prime,
\mu_{2}^\prime; \dots\vert M \vert \vec{v};
\vec{k}_1, \mu_1; \vec{k}_{2}, \mu_{2};\dots\rangle & &\nonumber\\
&&\hspace{-4.5cm}\propto\,{v^0\, \delta^3(\vec{v}^\prime - \vec{v})}\,
\langle \vec{k}_1^\prime, \mu_1; \vec{k}_{2}^\prime,
\mu_{2}^\prime; \dots\vert\vert M \vert\vert
\vec{k}_1, \mu_1; \vec{k}_{2}, \mu_{2};\dots\rangle\, ,
\end{eqnarray}
leading to the separation of overall and internal motion of the system.

\section{Cluster separability}
A central requirement for local relativistic quantum field theories is \lq\lq microscopic causality\rq\rq , i.e. the property that field operators at space-time points $x$ and $y$ should commute or anticommute, depending on whether they describe bosons or fermions, if these space-time points are space-like separated, i.e.
\begin{equation}
[\Psi(x),\Psi(y)]_{\pm}=0 \quad\hbox{for}\quad (x-y)^2<0\, .
\end{equation}
The crucial point here is that this must hold for arbitrarily small space-like distances. This condition requires an infinite number of degrees of freedom and can therefore not be satisfied in relativistic quantum mechanics with only a finite number of degrees of freedom. What replaces microscopic causality in the case of relativistic quantum mechanics is the physically more sensible requirement of \lq\lq macroscopic causality\rq\rq, or also often called \lq\lq cluster separability\rq\rq . It roughly means that subsystems of a quantum mechanical system should behave independently, if they are sufficiently space-like separated.

In order to phrase cluster separability in more mathematical terms, we start with an $N$-particle state $|\Phi\rangle$ with wave function $\phi(\vec{p}_1,\vec{p}_2,\dots,\vec{p}_N)$ and decompose this $N$-particle system into two subclusters $(A)$ and $(B)$. Next one introduces a separation operator $U^{(A)(B)}_\sigma$ with the property that
\begin{equation}
\lim_{\sigma\rightarrow\infty} \langle\Phi\vert U^{(A)(B)}_\sigma\vert \Phi\rangle =0\, .
\end{equation}
The role of the separation operator will become clearer by means of an example. Let us consider (space-like) separation by a canonical boost. In this case subsystem $(A)$ is boosted with velocity $\vec{v}$ and subsystem $(B)$ with velocity $-\vec{v}$. The action on the wave function is then
\begin{equation}
\Bigl(U^{(A)(B)}_{\vec{v}}\phi\Bigr)(\vec{p}_{i\in (A)},\vec{p}_{j\in (B)})=\phi\Bigl(\overrightarrow{B({ -}\vec{v})p}_{i\in(A)},\overrightarrow{B({ \vec{v})p}}_{j\in(B)}\Bigr)
\end{equation}
and one has to consider the limit $\sigma=|\vec{v}|\rightarrow \infty$ in Eq.~(7).

Having introduced a separation operator we are now able to formulate cluster separability in a more formal way. In the literature one can find different notions of it. A comparably weak, but physically plausible requirement, is cluster separability of the scattering operator:
\begin{equation}
\mathrm{s-}\!\!\lim_{\sigma\rightarrow\infty} {U_\sigma^{(A)(B)}}^\dag\, S\, {U_\sigma^{(A)(B)}} = S^{(A)}\otimes S^{(B)}\, .
\end{equation}
It means that the scattering operator should factorize into the scattering operators of the subsystems after separation. For three-particle systems it has been demonstrated that this type of cluster separability can be achieved by a BT construction~\cite{Coester:65}.

A stronger requirement is that the  Poincar\'e generators become additive, when the clusters are separated. In a weaker version this means for the four-momentum operator that
\begin{equation}
\lim_{\sigma\rightarrow \infty}\langle \Phi \vert {U_\sigma^{(A)(B)}}^\dag \Bigl(P^\mu-P^\mu_{(A)}\otimes I_{(B)}-I_{(A)}\otimes P^\mu_{(B)}\Bigr)\, U_\sigma^{(A)(B)} \vert\Phi\rangle = 0\,,
\end{equation}
the stronger version is that
\begin{equation}
\lim_{\sigma\rightarrow \infty}\Bigl\vert\Bigl\vert\Bigl(P^\mu-P^\mu_{(A)}\otimes I_{(B)}-I_{(A)}\otimes P^\mu_{(B)}\Bigr)\, U_\sigma^{(A)(B)} \vert\Phi\rangle\Bigr\vert\Bigr\vert=0\, .
\end{equation}
The BT construction violates both conditions already in the $2$+$1$-body case (i.e. particles 1 and 2 interacting and particle 3 free)~\cite{Mutze:78,Kei:91}. The reason for the failure can essentially be traced back in this case to the fact that the BT-type mass operator and the mass operator of the separated 2+1-particle system differ in the velocity conserving delta functions. In the BT-case it is the overall three-particle velocity which is conserved, in the separated case it is rather the velocity of the interacting two-particle system. The separation, however, is done by boosting with the velocity of the interacting two-particle system.

One may now ask, whether wrong cluster properties lead to observable physical consequences. From our studies of the electromagnetic structure of mesons we have to conclude that this is indeed the case~\cite{Biernat:2009my,GomezRocha:2012zd,Biernat:2014dea}. In these papers electron scattering off a confined quark-antiquark pair was treated within relativistic point form quantum mechanics starting from a BT-type mass operator in which the dynamics of the photon is also fully included. The meson current can then be identified in a unique way from the resulting one-photon-exchange amplitude which has the usual structure, i.e. electron current contracted with the meson current and multiplied with the covariant photon propagator. The covariant analysis of the resulting meson current, however, reveals that it exhibits some unphysical features which most likely can be ascribed to wrong cluster properties. For pseudoscalar mesons, e.g., its complete covariant decomposition takes on the form
\begin{equation}
{\tilde{J}^\mu(\vec{p}_M^\prime;\vec{p}_M)} = {
(p_M+p^\prime_M)^\mu \, f(Q^2,{ s})}+ { (p_e+p^\prime_e)^\mu
\, g(Q^2,s)}\, .
\end{equation}
It is still conserved, transforms like a four-vector, but exhibits an unphysical dependence on the electron momenta which manifests itself in form of an additional covariant (and corresponding form factor) and a spurious Mandelstam-$s$ dependence of the form factors. Although unphysical, these features do not spoil the relativistic invariance of the electron-meson scattering amplitude. The Mandelstam-$s$ dependence of the physical and spurious form factors $f$ and $g$ is shown in Fig.~1. Since the spurious form factor $g$ is seen to vanish for large $s$ and the $s$-dependence of the physical form factor $f$ becomes also negligible in this case it is suggestive to extract the physical form factor in the limit $s\rightarrow\infty$. This strategy was pursued in Refs.~\cite{Biernat:2009my,GomezRocha:2012zd,Biernat:2014dea} where it lead to sensible results. It gives a simple analytical expression for the physical form factor $F(Q^2)=\lim_{s\rightarrow\infty} f(Q^2,s)$ which agrees with corresponding front form calculations in the $q_\perp=0$ frame. Similar effects of wrong cluster properties on electromagnetic form factors were also observed in model calculations done within the framework of front form quantum mechanics~\cite{Keister:2011ie}.
\begin{figure}[h!]
\begin{center}
\includegraphics[width=6cm]{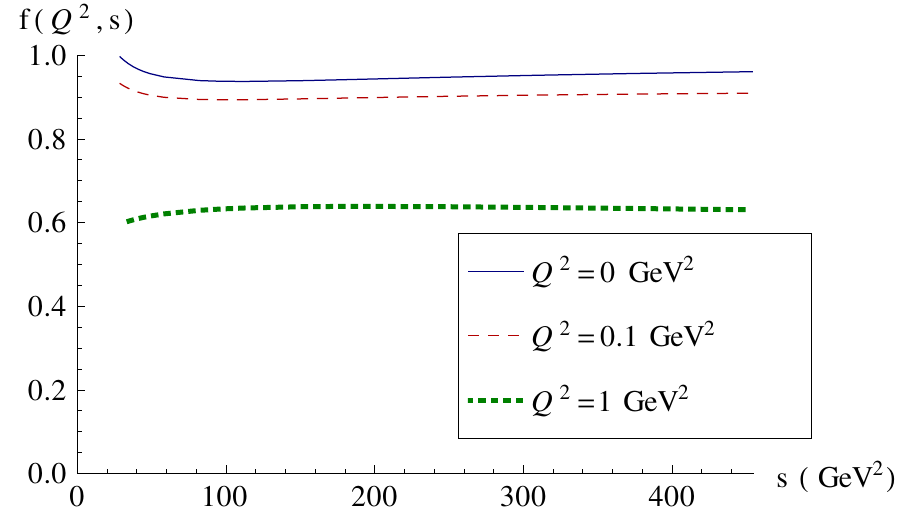}\hspace{2.0cm}
\includegraphics[width=6cm]{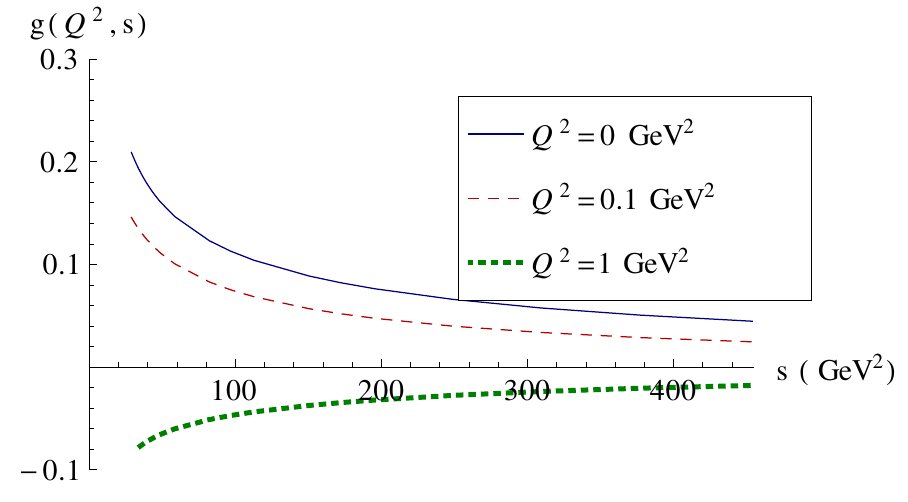}
\caption{Mandelstam-$s$ dependence of the physical and spurious $B$ meson electromagnetic form factors $f$ and $g$ for various values of the (negative) squared four-momentum transfer $Q^2$~\cite{GomezRocha:2012zd}. The result has been obtained with a harmonic-oscillator wave function with parameters $a=0.55$~GeV, $m_b=4.8$~GeV, $m_{u,d}=0.25$~GeV.}
\end{center}\label{figure1}
\end{figure}

\section{Restoring cluster separability}
It is obviously the BT-type structure of the four-momentum operator (see Eq.~(\ref{eq:BT})) which guarantees Poincar\'e invariance on the one hand, but leads to wrong cluster properties on the other hand (if one has more than two particles). In order to show, how this conflict may be resolved, let us consider a three-particle system with pairwise two-particle interactions. To simplify matters we will consider spinless particles and neglect internal quantum numbers. We start with the four-momentum operators of the two-particle subsystems,
\begin{equation}
P^\mu_{(ij)}=M_{(ij)} V^\mu_{(ij)}\,,\quad i,j=1,2,3\, ,\quad i\neq j\, ,
\end{equation}
which have a BT-type structure (i.e. $V^\mu_{(ij)}$ is free of interactions). Cluster separability holds for these subsystems, if the two-particle interaction is sufficiently short ranged. The third particle can now be added by means of the usual tensor-product construction
\begin{equation}
\tilde{P}^\mu_{(ij)(k)}=P^\mu_{(ij)}\otimes I_{(k)}+I_{(ij)}\otimes P^\mu_{(k)}\,.
\end{equation}
The individual four-momentum operators
$\tilde{P}^\mu_{(ij)(k)}$ describe 2+1-body systems in a Poincar\'e invariant way and exhibit also the right cluster properties. One may now think of adding all these four momentum operators, to end up with a four momentum operator for a three particle system with pairwise interactions:
\begin{equation}
\tilde{P}^\mu_3=\tilde{P}^\mu_{(12)(3)}+\tilde{P}^\mu_{(23)(1)}+\tilde{P}^\mu_{(31)(2)}-2 {P}^\mu_{3\,\mathrm{free}}\, .
\end{equation}
But the components of the resulting four-momentum operator do not commute,
\begin{equation}
\bigl[\tilde{P}^\mu_3,\tilde{P}^\nu_3\bigr]{\neq} 0 \quad \hbox{since} \quad [M_{(ij)\,\mathrm{int}},V^\mu_{(j)}]\neq 0\, .
\end{equation}
One can, of course, write the individual $\tilde{P}^\mu_{(ij)(k)}$ in the form
\begin{equation}
\tilde{P}^\mu_{(ij)(k)}=\tilde{M}_{(ij)(k)}\, \tilde{V}^\mu_{(ij)(k)}\quad\hbox{with} \quad\tilde{M}_{(ij)(k)}^2=\tilde{P}_{(ij)(k)}\cdot \tilde{P}_{(ij)(k)}\, ,
\end{equation}
but the four-velocities $\tilde{V}^\mu_{(ij)(k)}$ contain interactions and differ for different clusterings, so that an overall four-velocity cannot be factored out of $\tilde{P}^\mu_3$. The key observation is now that all four-velocity operators have the same spectrum, namely $\mathbb{R}^3$. This implies that there exist unitary transformations which relate the four-velocity operators. One can find, in particular, unitary operators $U_{(ij)(k)}$ such that
\begin{equation}\label{eq:inter}
\tilde{V}^\mu_{(ij)(k)}=U_{(ij)(k)} V^\mu_3 U_{(ij)(k)}^\dag\, .
\end{equation}
With these unitary operators one can now define new three-particle momentum operators for a particular clustering,
\begin{equation}\label{eq:packing}
P^\mu_{(ij)(k)}:=U_{(ij)(k)}^\dag \tilde{P}^\mu_{(ij)(k)} U_{(ij)(k)}=U_{(ij)(k)}^\dag \tilde{M}_{(ij)(k)} U_{(ij)(k)} U_{(ij)(k)}^\dag \tilde{V}^\mu_{(ij)(k)} U_{(ij)(k)}= {M}_{(ij)(k)} V_3^\mu\, ,
\end{equation}
which have already BT-structure, i.e. with the free three-particle velocity factored out. From Eq.~(\ref{eq:packing}) it can be seen that the unitary operators $U_{(ij)(k)}$ obviously \lq\lq pack\rq\rq\ the interaction dependence of the four-velocity operators $\tilde{V}^\mu_{(ij)(k)}$ into the mass operator ${M}_{(ij)(k)}$. Therefore they were called \lq\lq packing operators\rq\rq\ by Sokolov in his seminal paper on the formal solution of the cluster problem~\cite{Sokolov:78}. The sum $({P}^\mu_{(12)(3)}+{P}^\mu_{(23)(1)}+{P}^\mu_{(31)(2)}-2 {P}^\mu_{3\,\mathrm{free}})$ describes a three-particle system with pairwise interactions, it has now BT-structure and satisfies thus the correct commutation relation. However, it still violates cluster separability. The solution is a further unitary transformation of the whole sum by means of $U=\prod U_{(ij)(k)}$, assuming that $U_{(ij)(k)}\rightarrow 1$ for separations $(ki)(j)$, $(jk)(i)$ and $(i)(j)(k)$. The final expression for the three-particle four-momentum operator, that has all the properties it should have, is:
\begin{eqnarray}
P^\mu_3 &:=&{U}\,\left[P^\mu_{(12)(3)} +P^\mu_{(23)(1)} +P^\mu_{(31)(2)}+{ P^\mu_{(123)\,\mathrm{int}}}-2P^\mu_{3\,\mathrm{free}}\right]\,{U^{\dagger}}\nonumber \\
&=&{ U}\left[ M_{(12)(3)}+ M_{(12)(3)}+ M_{(12)(3)}+{ M_{(123)\,\mathrm{int}}}-2 M_{3\,\mathrm{free}}\right]\,V_3^\mu\,{ U^{\dagger}}\nonumber\\&=&{ U}\,M_3\,V_3^\mu\,{ U^{\dagger}}\, .
\end{eqnarray}
If $U$ commutes with Lorentz transformations, it can be shown that such a "generalized BT construction" will satisfy relativity and cluster separability for $N$-particle systems.
In addition to the three-body force induced by $U$, which is of purely kinematical origin, we have also allowed for a genuine three-body interaction ${M_{(123)\,\mathrm{int}}}$. Since the $U_{(ij)(k)}$ will, in general, not commute, $U$ depends on the order of the $U_{(ij)(k)}$ in the product. For identical particles one should even take some kind of symmetrized product, for which also different possibilities exist~\cite{Sokolov:78,Kei:91}. This means that $P^\mu_3$ is, apart of the newly introduced free-body interaction ${M_{(123)\,\mathrm{int}}}$, not uniquely determined by the two-body momentum operators $P^\mu_{(ij)}$. There are even different ways to construct the packing operators $U_{(ij)(k)}$. All the unitary transformations leave, however, the on-shell data (binding energies, scattering phase shifts, etc.) of the two-particle subsystems untouched, they only affect their off-shell behavior.

The kind of procedure just outlined formally solves the cluster problem for three-body systems. Generalizations to $N>3$ particles and particle production have also been  considered~\cite{Coester:82}. Its practical applicability, however, depends strongly on the capability to calculate the packing operators for a particular system. A possible procedure can also be found in Sokolov's paper. The trick is to split the packing operator further
\begin{equation}
U_{(ij)(k)}=W^\dag(M_{(ij)}) W(M_{(ij)\,\mathrm{free}})
\end{equation}
into a product of unitary operators which depend on the corresponding two-particle mass operators in a way to be determined. With this splitting one can rewrite Eq.~(\ref{eq:inter}) in the form
\begin{equation}
 W(M_{(ij)\,\mathrm{free}}) V^\mu_3 W^\dag(M_{(ij)\,\mathrm{free}})=W(M_{(ij)})\tilde{V}^\mu_{(ij)(k)}W^\dag(M_{(ij)})\, .
\end{equation}
Since this equation should hold for any interaction the right- and left-hand sides can be chosen to equal some simple four-velocity operator, for which $V_{(ij)}^\mu\otimes I_k$ is a good choice. In order to compute the action of $W$ it is then convenient to take bases in which matrix elements of $V^\mu_3$, $V_{(ij)}^\mu\otimes I_k$ and $\tilde{V}^\mu_{(ij)(k)}$ can be calculated. This is the basis of (mixed) velocity eigenstates
\begin{equation}
|\vec{v}_{(12)};\vec{\tilde{k}}_1,\vec{\tilde{k}}_2,\vec{p}_3\rangle=
|\vec{v}_{(12)};\vec{\tilde{k}}_1,\vec{\tilde{k}}_2\rangle\otimes|\vec{p}_3\rangle
\end{equation}
of $M_{(ij)(k)\,\mathrm{free}}$ if one wants to calculate the action of $W(M_{(ij)\,\mathrm{free}})$ and corresponding eigenstates of $M_{(ij)(k)}$ if one wants to calculate the action of $W(M_{(ij)})$. It turns out that the effect of these operators is mainly to give the two-particle subsystem $(ij)$ the velocity $v_{(ij)(k)}$ of the whole three-particle system. After some calculations one finds out that the whole effect of the packing operator $U_{(ij)(k)}$ on the mass operator $\tilde{M}_{(ij)(k)}$ is just the replacement
\begin{equation}
\frac{1}{m_{(ij)}^{\prime\, 3/2} m_{(ij)}^{3/2}}
\, v^0_{(ij)}\delta^3(\vec{v}_{(ij)}^{\,\,\prime}-\vec{v}_{(ij)}) \rightarrow
\frac{\sqrt{v_{(ij)}^{\,\,\prime} \cdot {v}_{(ij)(k)}}}{m_{(ij)(k)}^{\prime\, 3/2}}
\frac{\sqrt{v_{(ij)} \cdot {v}_{(ij)(k)}}}{m_{(ij)(k)}^{3/2}}\, v^0_{(ij)(k)}\delta^3(\vec{v}_{(ij)(k)}^{\,\,\prime}-\vec{v}_{(ij)(k)})\, 
\end{equation}
in the mixed velocity-state matrix elements. Here $m_{(ij)}$ and $m_{(ij)(k)}$ are the invariant masses of the free two-particle subsystem and the free three-particle system, $v_{(ij)}$ and $v_{(ij)(k)}$ the corresponding four-velocities.

\section{Summary and outlook}
We have given a short introduction into the field of relativistic quantum mechanics. It has been shown that the Bakamjian-Thomas construction, the only known systematic procedure to implement interactions such that Poincar\'e invariance of a quantum mechanical system is guaranteed, leads to problems with cluster separability for systems of more than two particles. Cluster separability is a physically sensible requirement for quantum mechanical systems which replaces microcausality in relativistic quantum field theories. We have discussed the physical consequences of wrong cluster properties, e.g., unphysical contributions in electromagnetic currents of bound states. Following the work of Sokolov we have sketched how a three-particle mass operator with pairwise interactions and correct cluster properties can be constructed. This is accomplished by a set of unitary transformations called packing operators. For the simplest case of three spinless particles we have explicitly calculated these packing operators. In a next step it is planned to use these results to see whether the problems encountered with electromagnetic bound-state currents can be cured by starting with a mass operator that has the correct cluster properties.

\end{document}